\documentstyle[aps,twocolumn]{revtex}

\begin{document}

\title{Segregation in a competing and evolving population}
\author{P. M. Hui$^{1}$, T. S. Lo$^{1}$, and N. F. Johnson$^{2}$}
\address{$^{1}$ Department of Physics, The Chinese University of Hong Kong
\\
Shatin, New Territories, Hong Kong \\
$^2$ Department of Physics, The University of Oxford \\ 
Parks Road, Oxford, OX1 3PU, United Kingdom}

\maketitle

\begin{abstract}
We study a recently proposed model in which
an odd number of agents are competing to be in the minority.  
In the model, the agents have one strategy in hand which is to
follow the most
recent history.  Each agent is also assigned a value $p$, which is
the probability that an agent will follow the trend.  Evolution 
is introduced through the modification of the value of $p$ when
the performance of an agent becomes unsatisfactory.  We present
numerical results for the distribution of $p$ values in the population
as well as the average duration between modifications at a given $p$ for
different values of the parameters in the model.  Agents who either
always follow the trend or always act opposite to the trend, tend
to out-perform the cautious agents.  We also point out
the difference between the present model and a slightly modified model
in which a strategy is randomly assigned to every agent initially.

\end{abstract}

\vspace*{0.4 true in}
\noindent {\small \em 
Paper presented at Dynamics Days Asia Pacific: First
International Conference on Nonlinear Science, July 13 - 16, 1999 at
the Hong Kong Baptist University, Hong Kong. 

\noindent Proceedings to be published in
Physica A.
}

\noindent{\small E-mail: pmhui@phy.cuhk.edu.hk}

\section{Introduction}

Agent-based models of complex adaptive systems (CAS) provide invaluable 
insight into the highly non-trivial global behaviour of a population of 
competing agents\cite{holland}.  Typically, these models involve agents 
with similar capability competing for a limited resource.  The agents 
share the same global information, which is in turn generated by 
the behaviour of the agents themselves, and they learn from past 
experience.  A realistic situation, for example, is where the index 
of a stock market is made known to every participating agent: the agents  
must then decide whether to buy or sell based on this global information. 
It is, therefore, not surprising to see that these models constitute 
an active part of the growing field of econophysics\cite{stanley,confs}. 

One of the earliest models was proposed by Arthur\cite{arthur}.  The model, 
which is referred to as the bar-attendance model, consists of $N$ agents 
trying to decide whether to attend a bar with a seating capacity 
$L$ ($L < N$).  The attendance in the past weeks is announced 
to all agents, forming the global information.  A correct decision 
is to attend (not to attend) the bar with an attendance less than or 
equal to (higher than) the cutoff $L$.  It turns out 
that\cite{arthur,johnson1} the population cooperates or self-organizes 
in the sense that the attendance in each turn is usually close to $L$ and
the
variance of the attendance can show a minimum  with increasing adaptability
of the
individual agents.  This model shows  a few features typical of CAS.  The
cooperation
in the system necessarily  requires an inhomogeneous population.  If all
agents were
to decide  according to the same strategy, all of them would act identically
and 
hence lose.  In fact, there is no {\em a priori} best strategy and 
a strategy good at one point in time will become bad when too many 
agents use it.  The agents hence interact through the creation and 
sharing of the global information, and are forced to make decisions 
based on inductive, rather than deductive thinking\cite{science}. 

The bar-attendance model is rather complicated in that the actual 
attendance is announced.  Challet and Zhang proposed a 
binary game, called the minority game, in which an odd number $N$ of agents 
are competing to be in the minority group\cite{challet,savit}.
The agents decide to go into
one of two rooms, with the winners being those in the room with fewer 
agents.  The outcomes are announced and form the global information.  The 
agents are assumed to have limited and similar capabilities in that they all
decide based on the outcomes of the recent $m$ turns.  There are a total 
of $2^{m}$ possible history bit-strings of length $m$, thus forming 
a strategy space consisting of a total of $2^{2^{m}}$ strategies.  Each 
agent picks $s$ strategies from this pool initially and uses the one with
the best 
accumulative performance in deciding the next move.  Detailed numerical 
calculations have revealed that the standard deviation (SD) in the
attendance 
in a room shows a minimum as a function of $m$ at a value corresponding 
to $N \cdot s \sim 2\cdot 2^{m}$.  Johnson and coworkers \cite{crowd} 
suggested that the features can be understood in terms of 
the dynamical formation of crowds consisting of agents using, say, the 
best strategy at a particular moment in time and anticrowds consisting 
of agents using the strategy anti-correlated to the crowd.  Basically 
the degree of overlap in the strategies among the agents plays 
an important role.  The condition $N\cdot s \sim 2\cdot 2^{m}$ 
favours the formation of crowds and anti-crowds of comparable size and 
leads to a minimum in the SD.  The idea can be formulated quantitatively 
in terms of the Hamming distance\cite{dHR} between strategies.  

In both the bar-attendance and the basic minority game, the strategies, 
once distributed at the beginning of the game, are fixed and the agents 
adapt by choosing the best-performing strategy in hand.  Recently, we 
proposed a simple model consisting of an {\em evolving} population in 
which the agents can adapt by changing their behaviour without the 
limitation imposed by the strategies being initially
distributed\cite{prl,royal}. 
In this paper, we report results on this model focusing on the 
possible segregation of the population as a result of competition and 
evolution.  We introduce the model in Sec.2.  Results are presented 
in Sec.3.  In Sec.4, we summarize our results and point out the 
difference between our model and a similar model proposed in 
Ref.\cite{dHR} in an attempt to formulate a theory of the present model. 

\section{The genetic model}

We introduce a simple, yet realistic, model for an evolving population 
containing adaptive agents who compete to be in the minority.  
Inspired by Ref.\cite{challet}, we consider the model of an odd number 
$N$ of agents repeatedly choosing to be in room ``0" or room ``1".  
After each agent has independently chosen a room, the winners are those 
in the minority room.  The ``output" for each time step is a single 
binary digit denoting the minority room.  Each agent is given 
the information of the most recent $m$ outcomes.  Each agent also has 
access to a common register or ``memory" containing the outcomes 
from the most recent occurrences of all $2^{m}$ possible bit strings 
of length $m$.
Consider, for example, $m=3$ and denote ($xyz$)$w$ as
the $m=3$ bit string ($xyz$) 
and outcome $w$.  An example memory would comprise (000)1, (001)0,
(010)0, (011)1, (100)0, (101)1, (110)0, (111)1.  Following a run of
three wins for room ``0" in the recent past, the winning room was subsequently
``1".  
Faced with a given bit string of length $m$,
it seems reasonable for an agent to simply predict the same outcome as that 
registered in the memory.
The agent will hence choose room ``1" following the next 000 sequence.  If
``0" turns out to be the winning room, the entry (000)1 in the memory
is then updated to be (000)0. 
Simply put, each agent looks into the most recent
history for the same pattern of $m$ bit string and predicts the 
outcome using the history.  In effect, each agent holds one strategy 
and all agents hold the same strategy, with the strategy
being {\em dynamical}.
The strategy is hence to follow the trend.
However, if all $N$ agents act
in the 
same way, they will all lose.  A successful agent is one who 
can follow a trend as long as it is valid and to correctly predict when
it will end.  To incorporate this factor into our model, we assign 
to each agent a single number $p$, which we refer to as 
the ``gene"-value.  Following a given $m$-bit sequence, $p$ is the 
probability that the agent will choose the same outcome as that stored 
in the memory, i.e., he will follow the current predictor.  An 
agent will reject the prediction and choose the opposite 
action with probability $1-p$.  

To incorporate evolution into our model, we assign $+1$ ($-1$) point to 
every agent in the minority (majority) room at each time step.  If an 
agent's score falls below a value $d$ ($d<0$), his gene value 
is modified.  The new $p$ value is chosen randomly from a range of values 
centered on the old $p$ with a width equal to $R$.  We impose 
reflective boundary condition to ensure that $0\leq p \leq 1$.  Our 
conclusions do not depend on the particular choice of boundary 
conditions.  For $R=0$, the agents will not change their gene values 
at all.  For $R=2$, the new gene value is uncorrelated with the old one
upon modification.  

\section{Results}

We have carried out detailed numerical studies of our model. 
Initially, each agent is randomly assigned a 
gene value in the range $0 \leq p \leq 1$. The population 
is allowed to evolve.  We focus on two quantities, $P(p)$ and $L(p)$, 
in the asymptotic limit.  Here $P(p)$ is the frequency distribution 
of gene values, typically taken in the long time limit over a time 
window and normalized to unity; $L(p)$ is the lifespan defined as 
the average length of time a gene 
value $p$ survives between modifications.  Figure 1 shows $L(p)$ and 
$P(p)$ (inset) as a function of 
$p$ for different values of $m$.  
The other parameters are taken to be $N=101$, $R=0.2$ and $d=-4$.  The most 
interesting feature is that $P(p)$ becomes peaked around $p=0$ and 
$p=1$, with a similar behaviour in $L(p)$.  The results are insensitive 
to the initial distribution of $p$.  Surprisingly the results indicate 
that agents who either always follow or never follow what happened last 
time, generally perform better than cautious agents using 
an intermediate value of $p$.  Figure 1 also shows that there is no explicit
dependence on $m$ for $P(p)$ and $L(p)$.  We have also checked that 
different values of $d$ do not change the normalized distribution $P(p)$.  
The lifespan $L(p)$ obviously does depend on $d$ as shown in Fig. 2 for 
$d = -1, -2, \cdots, -9$.  A more negative value of $d$ leads to a longer 
time between modifications and hence a longer time in approaching 
the asymptotic $P(p)$.  The inset in Fig. 2 shows $L(p)/|d|$ as a
function of $p$ for different values of $d$ and all the data 
fall onto one curve showing $L(p) \sim |d|$.  

Figure 3 shows the dependence of $L(p)$ and $P(p)$ (inset) as a function of
$p$ for different values of $R$.  Different values of $R$ affect the
time it takes to approach the asymptotic limit, since a larger value of $R$
gives the agent a larger jump in gene-value space in order to arrive at the
final $P(p)$.  However, $L(p)$ and $P(p)$ do not depend on the value of $R$.
It should be pointed out that even when $P(p)$, and hence $L(p)$, takes
on its asymptotic form, agents are still constantly modifying their gene
values.
It is a dynamical state in that agents are changing their gene values while
keeping the form of $P(p)$ unchanged.  Figure 4 shows $P(p)$ and $L(p)$
for games with different numbers of agents.  The normalized $P(p)$, again,
does not depend on $N$, while the $L(p)$ is generally higher for
games with larger $N$.  Our results thus show that the segregation in the
population as indicated in $P(p)$ is robust and insensitive to the choice
of parameters in our model.  It merely comes from the desire of
the agents to do the opposite of the majority.
The segregation implies that the population as a whole samples the 
microstates of the system, 
i.e. the different possible distributions of the gene
value in the population, unevenly as time evolves.  
There are microstates 
in which the total points deducted per turn are relatively small and
the population tends to stay in these microstates longer 
giving rise to the segregation\cite{prl}.

Note also that
the results are symmetrical about $p=1/2$ for both $P(p)$ and $L(p)$.
Basically,
it follows from a symmetry in the game in that the past behaviour
contains the same information for every agent, and hence there is no
advantage to any agent.  Therefore, the game has the same distribution
$P(p)$ regardless of whether the real, or static, or random history
is followed; and we could replace the strategy at any moment by its inverse.
Thus, agents with gene values $p$ and $1-p$ are doing equally well.

\section{Summary}
We have presented numerical results for our genetic 
model consisting of an
evolving and competing population.  Agents who
always follow or always act opposite to the predictor, out-perform
the cautious agents.  Recently, an attempt was made to explain
quantitatively the form of the asymptotic gene distribution $P(p)$\cite{dHR}.
The theory, while succeeding in obtaining a $P(p)$ similar to those
reported here, was formulated based on a slightly different model.
In the modified model, 
one strategy in the pool of $2^{2^{m}}$
possible strategies (corresponding to a game with 
memory $m$) is assigned randomly
to each player at the beginning of the game, in addition to the gene value. 
Thus, the modified model in Ref.\cite{dHR} corresponds to a 
minority game with $s=1$ plus the inclusion of a gene value $p$ for 
each agent.  
Although the modified
model and the present model both correspond to cases
with one strategy per agent, the strategy is dynamical in the present model
and
is constantly updated; in the modified model the strategy is fixed. 
It is, therefore, interesting to check numerically if the two models give
identical
results for $P(p)$. Figure 5 shows $P(p)$ for the modified model with 
$m=1,2,\cdots,10$.  
We notice that, unlike our present model, the modified model's 
$P(p)$ depends on
$m$ with the small $m$ limit approaching the result of our model.  
The difference between the two models comes from the fact that for 
large values of $m$, the strategies held by the agents in the modified 
model are likely to be uncorrelated.  In this case, each 
agent cannot adapt to the behaviour of the other agents and the 
distribution $P(p)$ becomes flatter as $m$ increases.  
Interestingly,
the quantitative analysis in Ref.\cite{dHR}, which is based on the
modified model, does not give an $m$-dependent $P(p)$.  

A more complete
theory of our present genetic model can be formulated by investigating the
attendance distribution in one of the two rooms.
The approach\cite{unpublished} is
to relate the average success rate of an agent in an $N$-agent game to a
($N-1$)-agent game with a particular agent being singled out, and to
derive an expression for the average success rate in which the
effect of the complicated interaction (and self-interaction) of
the agents is isolated.
Such a consideration leads to an average success rate $\tau(p)$ 
for an agent using a gene value $p$ of the form $\tau(p) \sim 1/2 -
A(N) p (1-p)$, where $A$ is an $N$-dependent
parameter which decreases with $N$.  The lifespan $L(p)$
is related to the average success rate by $L(p) \sim d/(1/2 - \tau(p))$
\cite{dHR}, hence
leading to the symmetry about $p=1/2$ for both $L(p)$ and $P(p)$ as
discussed in the last section \cite{lhtang}.  
Results along this line will be reported
elsewhere\cite{unpublished}.

Our model forms the basis for incorporating various interesting
complications.
One possibility is that
the agents, instead of competing to be in the minority group, are trying
to attend a room with a specific cutoff capacity $L$; the winners are
those deciding to attend (not to attend) with the turnout being less than
or equal to (greater than) the cutoff capacity.  In this case, it is
possible
for the population distribution $P(p)$ to become frozen, i.e. no further
modifications
of gene values among the agents, as time evolves for large (or small) enough
value of the cutoff\cite{freezing}.

\begin{center}
{\bf Acknowledgments}
\end{center}

One of us (PMH) would like to thank Prof. Bambi Hu for his efforts in having
made the conference such a successful and memorable one.

\begin{figure}
\caption{The lifespan $L(p)$, which is the average duration between
modifications for a gene value $p$,
as a function of gene value $p$ for 
$m=1,2,\cdots,8$.  The inset shows the distribution of gene values $P(p)$
as a function of $p$ for different values of $m$.  Both $L(p)$ and
$P(p)$ are insensitive to $m$. The other
parameters are $N=101$, $d=-4$ and $R=0.2$.}
\vspace*{0.3 true in}

\caption{The lifespan $L(p)$ for different values of $d$.  The curves
at $p=0.5$ from bottom to top correspond to $d=-1,-2,\cdots,-9$.  The inset
shows that all the data fall onto one curve if we plot $L(p)/(-d)$ as
a function of $p$. The other parameters are $N=101$, $m=3$ and $R=0.2$.}
\vspace*{0.3 true in}

\caption{The lifespan $L(p)$ and the distribution of gene values $P(p)$
(inset) as a function of $p$ for $R = 0.1,0.2,\cdots,2.0$.  Note that
both $L(p)$ and $P(p)$ are insensitive to $R$. The other parameters
are $N=101$, $m=3$ and $d=-4$.}
\vspace*{0.3 true in}

\caption{The lifespan $L(p)$ as a function of $p$ for $N=11,21,\cdots,81$.
The inset shows $P(p)$ for different values of $N$.  The other parameters
are $m=3$, $d=-4$ and $R=0.2$.}
\vspace*{0.3 true in}

\caption{The distribution of gene values $P(p)$ as a function of $p$
for the modified model in which each agent is assigned a strategy
initially.  At $p=0.5$, the curves from bottom to top correspond to
$m=1,2,\cdots,10$.  The $P(p)$ depends on $m$ in the modified model in
contrast to the genetic model.}
\end{figure}

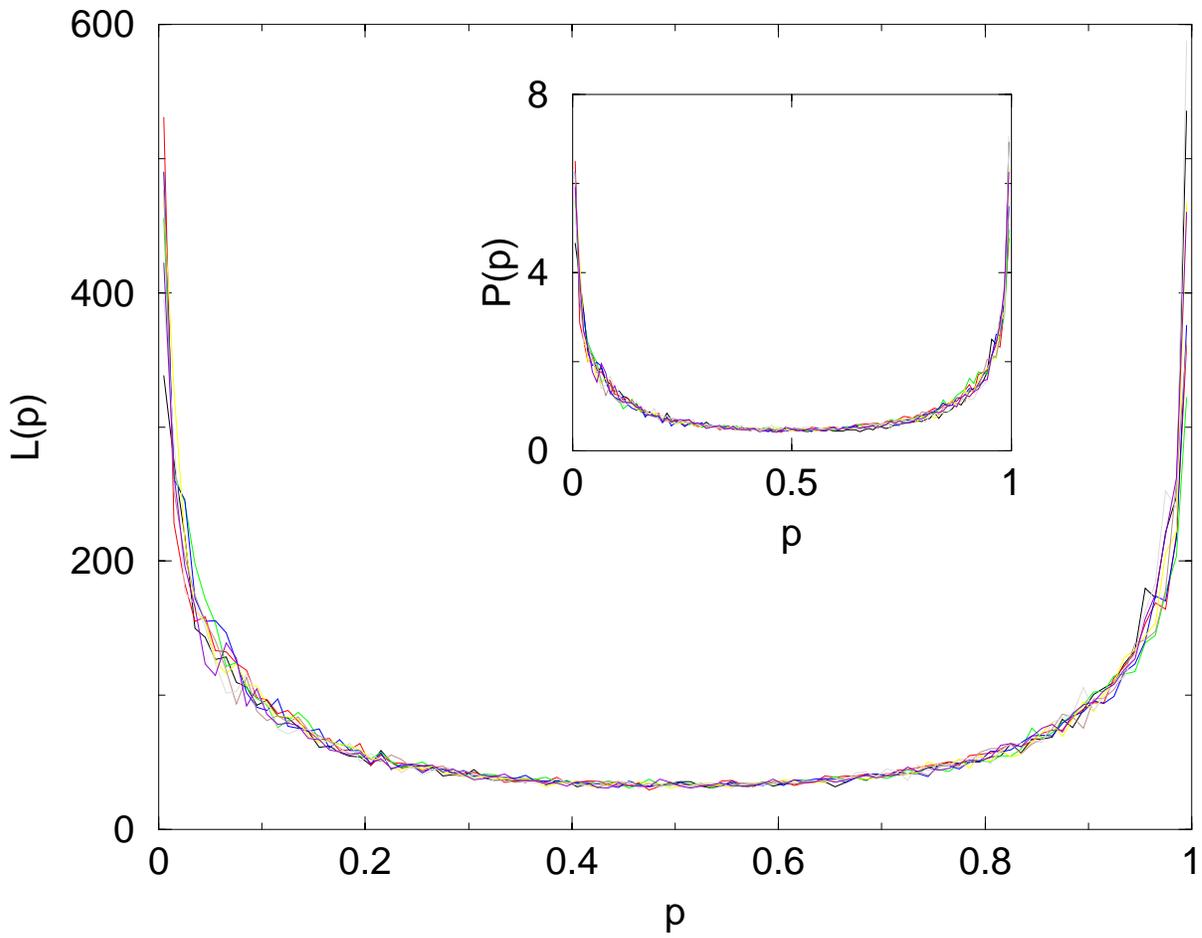
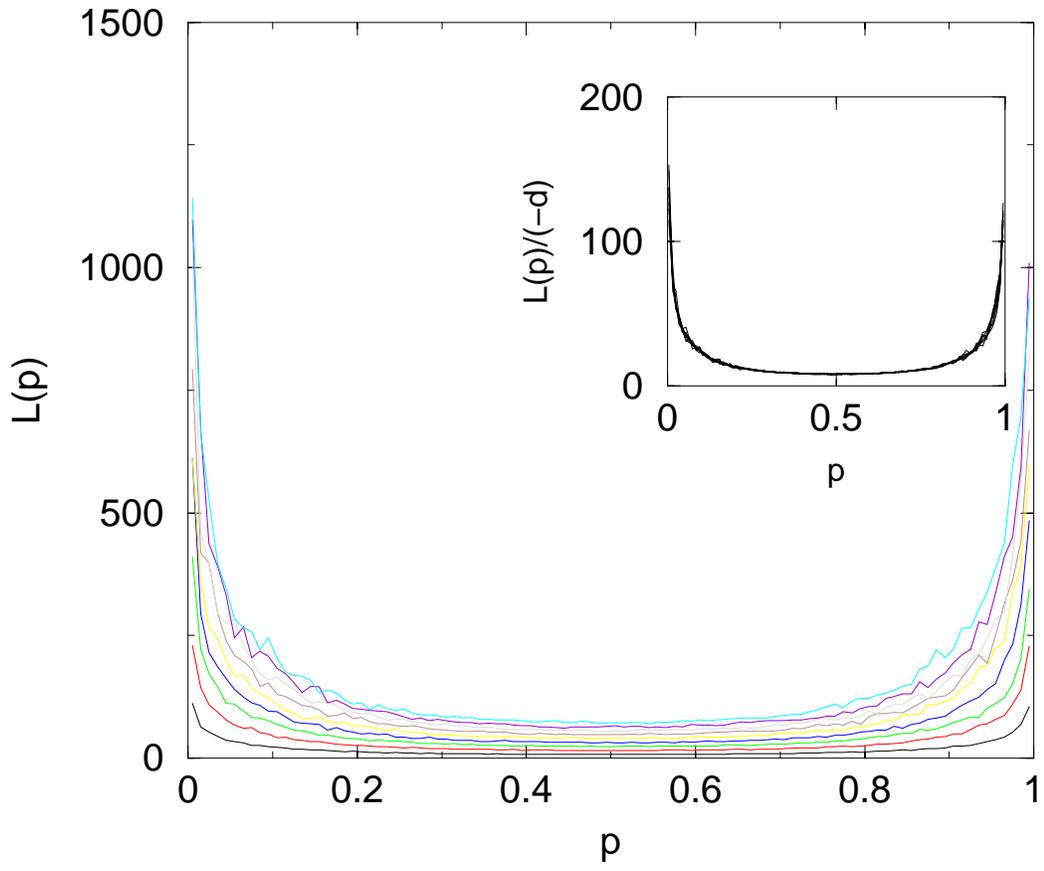
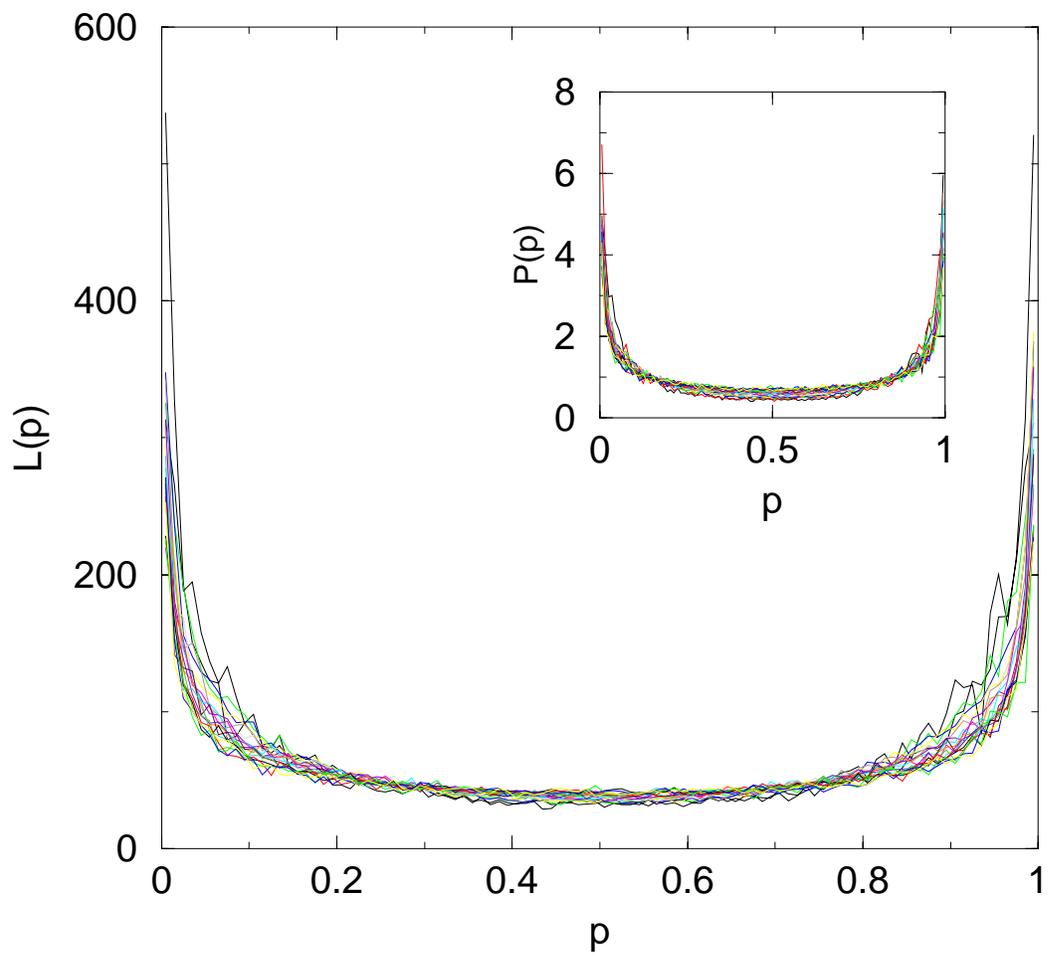
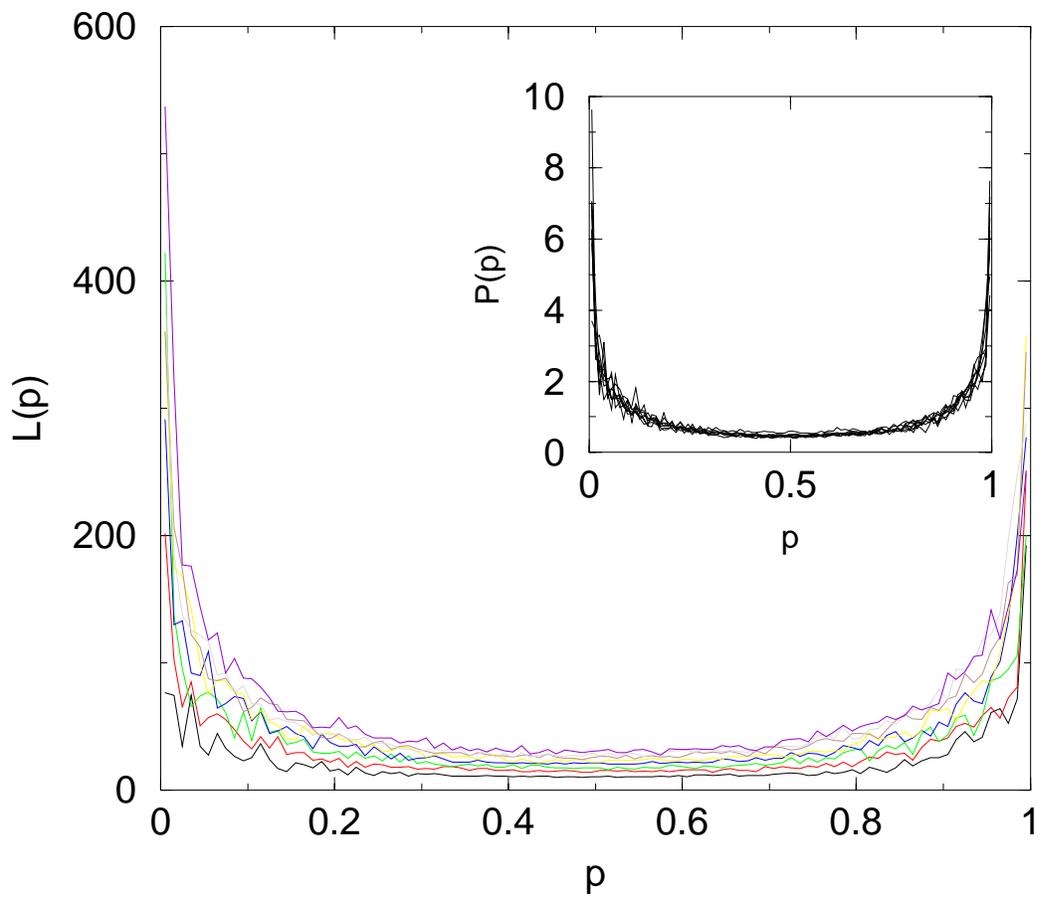
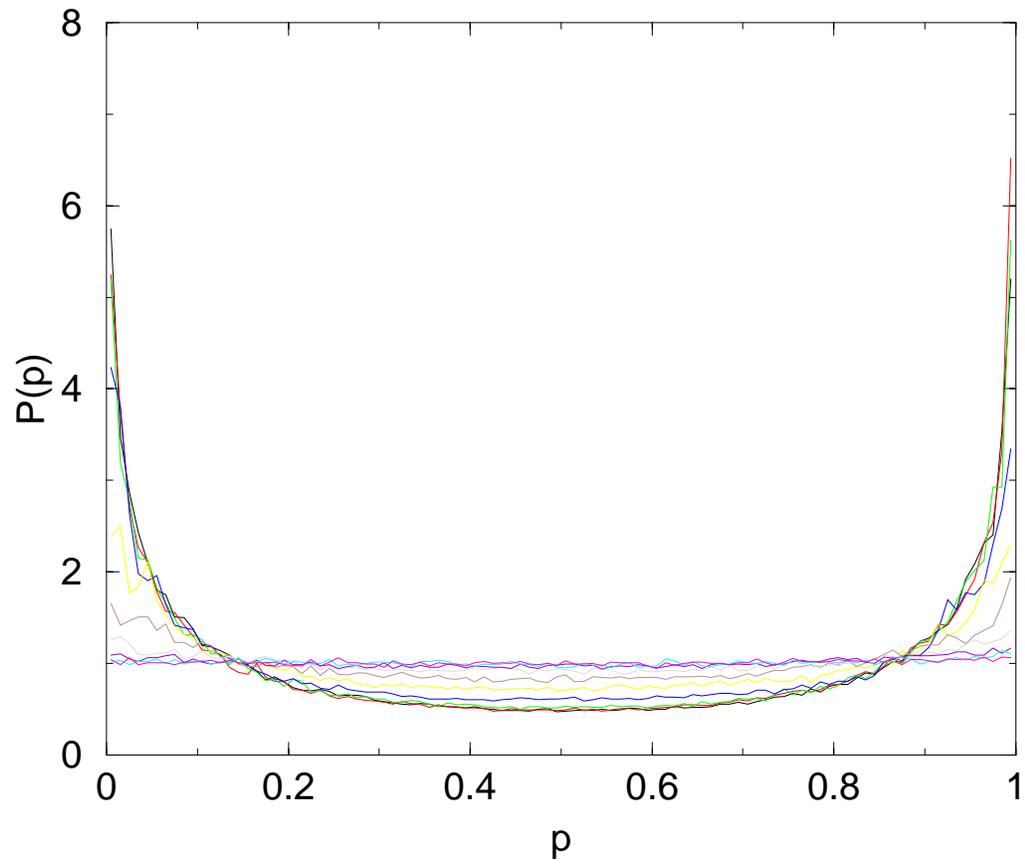
\end{document}